\documentclass[
reprint,
superscriptaddress,
amsmath,amssymb,
aps,prl
]{revtex4-2}

\usepackage[dvipsnames]{xcolor}
\usepackage{overpic}
\usepackage{tabularx}
\usepackage{graphicx}
\usepackage{dcolumn}
\usepackage{bm}
\usepackage{color}
\usepackage[unicode]{hyperref}
\hypersetup{
unicode=true,         
plainpages=false,
colorlinks=true,       
linkcolor=blue,          
citecolor=blue,        
urlcolor=blue           
}

\usepackage{soul}
\usepackage{pdfpages}

\newcommand{\m}[1]{\textcolor{black}{#1}}

\makeatletter
\AtBeginDocument{
    \pdfpagewidth=210mm
    \pdfpageheight=297mm
    \global\pdfpageattr\expandafter{\the\pdfpageattr/Rotate 0}
    \let\LS@rot\relax
}
\makeatother

\begin{document}

\preprint{APS/123-QED}

\title{\textbf{High-Resolution Spectroscopy of $^{173}$Yb$^{+}$ Ions} }

\author{J.~Jiang}
\email{jian.jiang@ptb.de}
\affiliation{Physikalisch-Technische Bundesanstalt, Bundesallee 100, 38116 Braunschweig, Germany}
\author{A.~V.~Viatkina}
\affiliation{Physikalisch-Technische Bundesanstalt, Bundesallee 100, 38116 Braunschweig, Germany}
\affiliation{Institut für Mathematische Physik, Technische Universität Braunschweig, Mendelssohnstraße 3, 38106 Braunschweig, Germany}
\author{Saaswath JK}
\affiliation{Physikalisch-Technische Bundesanstalt, Bundesallee 100, 38116 Braunschweig, Germany}
\author{M.~Steinel}
\affiliation{Physikalisch-Technische Bundesanstalt, Bundesallee 100, 38116 Braunschweig, Germany}
\author{M.~Filzinger}
\affiliation{Physikalisch-Technische Bundesanstalt, Bundesallee 100, 38116 Braunschweig, Germany}
\author{E.~Peik}
\affiliation{Physikalisch-Technische Bundesanstalt, Bundesallee 100, 38116 Braunschweig, Germany}
\author{S.~G.~Porsev}
\affiliation{Department of Physics and Astronomy, University of Delaware, Newark, Delaware 19716, USA}
\author{M.~S.~Safronova}
\affiliation{Department of Physics and Astronomy, University of Delaware, Newark, Delaware 19716, USA}
\author{A.~Surzyhkov}
\affiliation{Physikalisch-Technische Bundesanstalt, Bundesallee 100, 38116 Braunschweig, Germany}
\affiliation{Institut für Mathematische Physik, Technische Universität Braunschweig, Mendelssohnstraße 3, 38106 Braunschweig, Germany}
\affiliation{Laboratory for Emerging Nanometrology, Langer Kamp 6a/b, 38106 Braunschweig, Germany}
\author{N.~Huntemann}
\email{nils.huntemann@ptb.de}
\affiliation{Physikalisch-Technische Bundesanstalt, Bundesallee 100, 38116 Braunschweig, Germany}


\date{\today}

\begin{abstract}
Compared to other stable isotopes of $\rm{Yb}^+$, $^{173}$Yb$^+$ has a richer hyperfine structure, which leads to more favorable clock transitions, spectroscopic techniques for probing new physics, and more sophisticated quantum computing architectures.
However, to date, its electronic spectrum remains poorly characterized. Here, we report on efficient laser cooling, state preparation, and detection of a single trapped $^{173}\rm{Yb}^+$ ion. The previously unobserved $^2\!S_{1/2} \rightarrow {}^2\!D_{3/2}$ electric quadrupole transition at 436 nm is coherently excited, and the isotope shift between $^{171}\rm{Yb}^+$ and $^{173}\rm{Yb}^+$ on this transition is determined with an uncertainty of 1.4 Hz. Using microwave spectroscopy, we resolve the hyperfine structure (HFS) of the ${}^2\!D_{3/2}$ state with a relative uncertainty below $10^{-8}$. 
From the HFS measurement data, we infer for ${}^{173}$Yb a nuclear magnetic octupole moment $\Omega = -0.062(8)\,({\rm b} \times \mu_N)$ with uncertainty reduced by more than 2 orders of magnitude compared to previous studies.
The data also allow us to determine hyperfine anomalies for the ${}^2\!S_{1/2}$ and ${}^2\!D_{3/2}$ states. 
\end{abstract}

\maketitle


Trapped $\rm{Yb}^+$ ions of different isotopes have been used in various fundamental research and practical applications. Isotope-shift spectroscopy of $\rm{Yb}^+$ ions with zero nuclear spin has been employed in search of new bosons and to study higher-order atomic and nuclear effects \cite{Counts2020,Hur2022,Door2025}. The $^{171}\rm{Yb}^+$ ions with a nuclear spin of 1/2 were utilized to build accurate optical clocks based on the $^2\!S_{1/2} \rightarrow {}^2\!D_{3/2}$ \cite{Tamm2014, STUHLER2021} and ${}^2\!S_{1/2} \rightarrow {}^2\!F_{7/2}$
\cite{Huntemann2016,Tofful2024} transitions, which provide first-order magnetic field immune transitions between $m_F=0$ Zeeman levels. These clocks have been employed to search for violations of local Lorentz symmetry \cite{Sanner2019}, temporal variations in the fine structure constant \cite{Lange2021}, and ultralight dark matter \cite{Filzinger2023}. The $^{171}\rm{Yb}^+$ ions have also been chosen for building quantum computers motivated by the technical advantages over other ion species \cite{Nop2021, Wright2019}.

Compared with singly ionized ytterbium of other stable isotopes, 
$^{173}$Yb$^+$ is a promising candidate to extend existing research due to its deformed nucleus and large nuclear spin of 5/2.
\m{This includes studies of nuclear-spin-dependent parity-nonconservation (PNC) interactions between nuclei and electrons using the $^2\!S_{1/2}\rightarrow {}^{2}\!D_{3/2}$ transition \cite{Dzuba2011};} more accurate optical clocks based on the $^2\!S_{1/2} \rightarrow {}^2\!F_{7/2}$ transition with a reduced probe light shift resulting from a stronger coupling between the ground and excited state, as predicted in Ref. \cite{Dzuba2016}; investigations of nuclear deformation and higher-order nuclear moments by measuring the abundant hyperfine structure (HFS) of states with large angular momenta\cite{Xiao2020}; and qudit-based quantum information architectures \cite{Allcock2021,Jain2024}. Furthermore, differential studies between $^{173}$Yb$^+$ and $^{171}$Yb$^+$ ions, such as the PNC ratio \cite{Sahoo2011, Dzuba2011} and the hyperfine anomaly \cite{Roberts2021}, can profit from the similarity of their electronic structures, which significantly reduces the theoretical effort required to predict energy differences with high precision. 

\begin{figure}[h]
\centering
\includegraphics[trim={0cm 0cm 0cm 0cm},clip,width=0.48\textwidth]{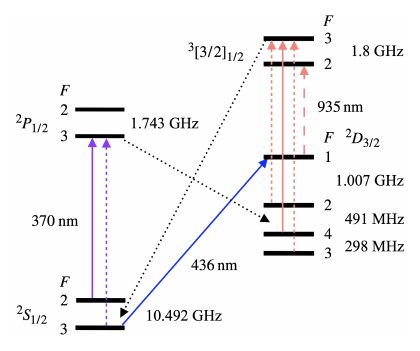}
\caption{Reduced energy level diagram of $^{173}$Yb$^+$ (not to scale). Laser cooling is realized on the $^2\!S_{1/2} \rightarrow {}^2\!P_{1/2}$ transition at 370 nm. Interruptions of the cooling cycle resulting from decay to the $^2\!D_{3/2}$ state are prevented using the repumping transition at 935 nm. Dotted lines indicate spontaneous decay, solid lines represent applied laser radiation, and dashed lines show additional spectral components of the laser radiation generated using electro-optic modulators. \m{The component indicated by the long-dashed line is responsible for the depopulation of the $^2\!D_{3/2}(F=1)$ level and is thus deactivated during state detection.} 
}
\label{fig:1}
\end{figure}

Although promising, the investigation of $^{173}$Yb$^+$ ions has so far been hampered by its more complex hyperfine structure (see Fig.~\ref{fig:1}).
Experiments with this isotope have resulted in a measurement of the $^2\!S_{1/2}$ ground state HFS using laser-microwave double resonance spectroscopy of buffer-gas-cooled trapped ions \cite{Münch1987}, and demonstration of state preparation in the ${}^2\!F_{7/2}$ state for laser-cooled trapped ions via repeated excitation of the $^2\!S_{1/2} \rightarrow {}^2\!D_{5/2}$ transition and subsequent spontaneous decay \cite{Roman2021, Dellaert2025}.

In this Letter, laser cooling close to the Doppler limit of a single trapped $^{173}$Yb$^+$ ion is demonstrated using a scheme compatible with ${}^2\!D_{3/2}$ state detection. We utilize this scheme for coherent spectroscopy of the previously unobserved $^2\!S_{1/2} \rightarrow {}^2\!D_{3/2}$ electric quadrupole (E2) transition at 436 nm (see Fig. \ref{fig:1}). 
The isotope shift between $^{171}\rm{Yb}^+$ and $^{173}\rm{Yb}^+$ on this transition is determined with an uncertainty of 1.4 Hz. 
State preparation is implemented by using the E2 transition and state detection to project the ion to a desired quantum state. Further extending the employed methods, microwave (MW) spectroscopy with hertz-level uncertainties is performed to measure the HFS of the ${}^2\!S_{1/2}$ and ${}^2\!D_{3/2}$ states. The former serves as an in situ magnetic field sensor for the characterization of the second-order Zeeman effect in both measurements. From the measured HFS of the ${}^2\!D_{3/2}$ state, we infer a nuclear magnetic octupole moment for ${}^{173}$Yb with an uncertainty reduced by more than 2 orders of magnitude compared to previous studies. This measurement helps address recent debate \cite{Xiao2020} over the $^{173}\rm{Yb}$ nuclear octupole moment and also reveals a substantial differential hyperfine anomaly \cite{Bohr1950} in the ${}^2\!D_{3/2}$ state of Yb$^+$ ions.

Spectroscopy of laser-cooled $^{173}$Yb$^+$ ions has recently been experimentally demonstrated \cite{Roman2021,Dellaert2025}. \m{In contrast to these works, here we present a laser cooling scheme compatible with $^2\!D_{3/2}$ state detection, inspired by the corresponding study for $^{171}\rm{Yb}^+$ \cite{Tamm2000}.} Since the ${}^2\!D_{3/2}$ state is repumped during laser cooling to counteract branching from the ${}^2\!P_{1/2}$ state, the population of the ${}^2\!D_{3/2}$ state after excitation may not be detected by the absence of fluorescence when applying cooling laser radiation (see Fig. \ref{fig:1}).
To avoid unwanted depletion of the $^2\!D_{3/2}$ state, we consider electric dipole selection rules and choose ${}^2\!P_{1/2}(F=3)$ and ${}^3[3/2]_{1/2}(F=3)$ as the excited states of the cooling and repumping transitions, respectively. The former choice prevents the ion from decaying to ${}^2\!D_{3/2}(F=1)$ during laser cooling. The latter choice ensures that an ion originally in the ${}^2\!D_{3/2}(F=1)$ state will not be pumped back to the cooling cycle. Consequently, the population of this state can be detected as the absence of fluorescence (dark) while cooling laser radiation is applied.

For efficient laser cooling, all hyperfine levels that will be populated need to be addressed. \m{We steer the laser frequencies to resonantly excite the $^2\!S_{1/2}(F=3) \rightarrow {^2\!P_{1/2}}(F=2)$ (370\,nm) and $^2\!D_{3/2}(F=4) \rightarrow {^3[3/2]_{1/2}}(F=3)$ (935\,nm) transition.} MW signals are applied to electro-optic modulators (EOMs) for the generation of all other required spectral components.
An MW tone at 3.332 GHz is applied to the EOM for the 935 nm light to drive the $^2\!D_{3/2}(F=1) \rightarrow {}^3[3/2]_{1/2}(F=2)$ transition. \m{This is required to avoid long dark periods after rare off-resonant population during laser cooling or after successful $^2\!D_{3/2}(F=1)$ state detection for an intended excitation}. 
Therefore, this tone is deactivated during the 4~ms fluorescence collection for state detection, whose fidelity is limited by the lifetime of the
${}^2\!D_{3/2}(F=1)$ state (see Supplemental Material \cite{appendixA}\nocite{abdelhafiz2019}\nocite{Leibfried2003}\nocite{KozPorFla96}\nocite{KozPorSaf15}\nocite{RalKraRea11}\nocite{RobTayGat99}). Compared to the alternative approach of using ${}^2\!D_{3/2}(F=4)$ as the dark state, the presented method offers a higher intrinsic detection fidelity due to the larger splittings between the $F=1$ and other hyperfine levels of the ${}^2\!D_{3/2}$ state.

For spectroscopy, we repeat a basic interrogation sequence, which starts with Doppler cooling for 6~ms. Subsequently, the MW signal applied to the EOM for 370 nm laser radiation is turned off to prepare the ion in the $^2\!S_{1/2}(F=3)$ state. Further state preparation into a specific Zeeman level is discussed below. A single rectangular interrogation pulse is used to transfer the ion from the ground state to the excited state. Successful excitation is detected as discussed above after mapping either state to the $^2\!D_{3/2}(F=1)$ state using resonant excitation pulses. 

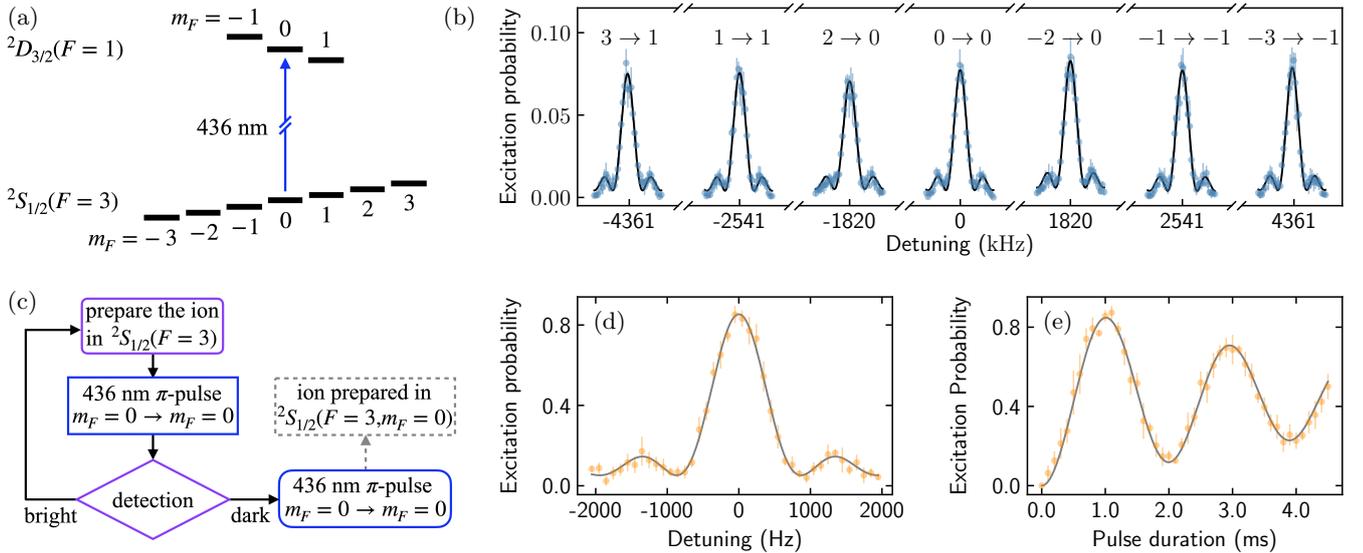
\begin{figure*}[htbp]
\centering
\begin{overpic}[trim={0cm 0cm 0cm 0cm},clip,width=\textwidth]{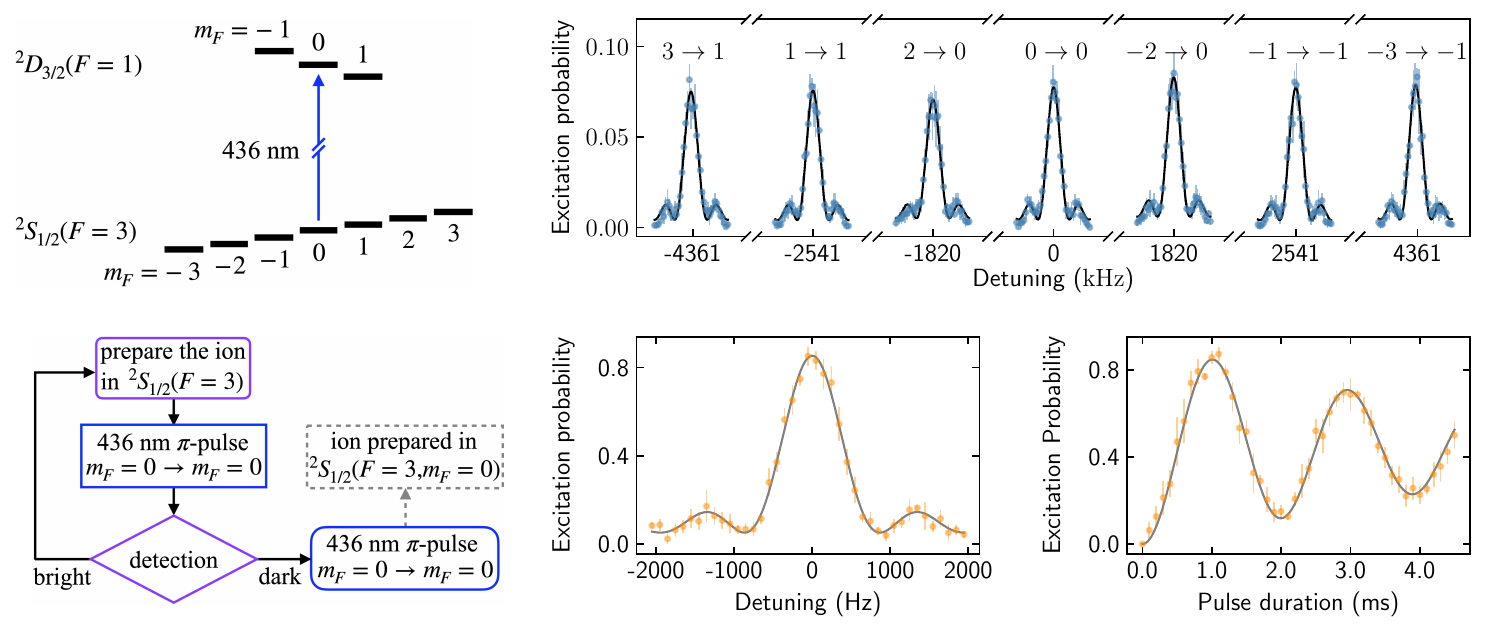}
    \put(1,40){(a)}
    \put(33,40){(b)}
    \put(1,19){(c)}
    \put(44,17.5){(d)}
    \put(77,17.5){(e)}
\end{overpic}
\caption{(a) Energy level diagram relevant for the $^2S_{1/2}(F=3) \rightarrow {}^2D_{3/2}(F=1)$ E2 transition with all Zeeman levels. (b) Spectra of $\pi$-pulse excitation on the E2 transitions from each Zeeman level of the $^2S_{1/2}(F=3)$ state. (c) Flow chart of the projective state preparation (PSP) using the E2 transition. Successful excitation of the 436~nm transition is indicated by the absence of fluorescence (dark) during detection. (d) Spectrum with $\pi$-pulse excitation and (e) observed Rabi flopping on the E2 transition between $m_F=0$ Zeeman levels using PSP. Deviations from full contrast are attributed to the ion temperature after Doppler cooling and the infidelity of the PSP process.}
\label{fig:3}
\end{figure*}

In experiments, the single $^{173}\rm{Yb}^+$ ion is confined using a Paul endcap trap \cite{abdelhafiz2019} operated under ultrahigh vacuum. The radial and axial trapping frequencies are $\omega_{x,y}/2\pi=598(5)$ kHz and $\omega_{z}/2\pi=1197(9)$ kHz, respectively. Ions are loaded from a resistively heated oven containing ytterbium metal via isotope-selective two-photon (399 and 370 nm) ionization \cite{Johaning2011} at the trap center. 
A constant magnetic field $\vec{B}$ ranging from 130~$\mu$T to 260~$\mu$T is applied perpendicular to the trap axis ($z$-axis) to destabilize coherent dark states during laser cooling \cite{Berkeland2002} and to provide a well-defined Zeeman shift during spectroscopy (see Supplemental Material \cite{appendixA} for details of the experimental setup). 
The 370 nm cooling beam, which overlaps with the 399 nm ionization beam, provides almost equal projection on all principal axes of the trap. Two additional 370 nm laser beams allow minimization of excess micromotion in all directions \cite{Keller2015}.
The 935 nm repumping beam, which is overlapped with the 436 nm E2 laser beam, lies in the plane spanned by the $z$-axis and $\vec{B}$ at an angle of about 45$^{\circ}$ to both directions. 
Fluorescence is collected on a photomultiplier tube for detection. The antennas used for MW spectroscopy are placed in front of the vacuum chamber along $\vec{B}$.
This system was initially operated with $^{171}\rm{Yb}^+$ as a single-ion optical clock, whose frequency shift effects have been investigated. A comparison with an independent $^{171}\rm{Yb}^+$ clock \cite{Huntemann2016} showed an agreement with $<10^{-16}$ uncertainty, which corresponds to $<100$~mHz for the 436~nm transition.

With this setup, we investigate the atomic structure of $^{173}\rm{Yb}^+$ ions by probing the $^2\!S_{1/2}(F=3) \rightarrow {}^2\!D_{3/2}(F=1)$ E2 transition at 436~nm (see Fig. \ref{fig:3}(a)) and employ this narrow transition in the preparation of a specific Zeeman state. 
The E2 probe laser is offset locked to that of a single $^{171}\rm{Yb}^+$ ion optical clock and inherits its $10^{-15}$ fractional short-term frequency instability \cite{Tamm2014, Huntemann2016}. 
The laser frequency can be adjusted via direct digital synthesis used in the generation of the offset frequency. 
The overlap between the probe laser beam and the trapped ion is optimized via the induced light shift determined with precision spectroscopy of the $^{171}\rm{Yb}^+$ ion. The polarization $\vec{E}_c$ and the wave vector $\vec{k}$ of the probe laser radiation, and the applied magnetic field $\vec{B}$ that defines the quantization axis, lie in one plane. The angle between $\vec{k}$ and $\vec{B}$, as well as between $\vec{E}_c$ and $\vec{B}$, is about $45^{\circ}$. This configuration allows excitations on the E2 transition with $\Delta m_F=0$ and $\Delta m_F=2$ and suppresses the $\Delta m_F=1$ component.

Figure~\ref{fig:3}(b) shows excitations of the E2 transition using 100-$\rm{}\mu$s-long $\pi$ pulses from each Zeeman level of the $^2\!S_{1/2}(F=3)$ ground state where the ion was initially prepared. Given about $10\%$ maximum excitation probability, we infer almost equal initial population in each Zeeman level of the $^2\!S_{1/2}(F=3)$ state. 
However, for high-contrast spectroscopy, preparing a specific single Zeeman level is advantageous. This is implemented with a technique first demonstrated in Ref.~\cite{Chou2017}. We repeatedly apply pulses of the 436 nm radiation 
and probe the population of the $^2\!D_{3/2}(F=1)$ state until it is populated. A single $\pi$ pulse then transfers the ion to the $^2\!S_{1/2}(F=3,m_F=0)$ state (see Fig. \ref{fig:3}(c) for a flow chart).  
Using this projective state preparation (PSP), Fig. \ref{fig:3}(d) shows the Fourier-limited spectrum obtained with 1 ms $\pi$ pulses on the $^2\!S_{1/2}(F=3,m_F=0) \rightarrow {}^2\!D_{3/2}(F=1,m_F=0)$ transition. By varying the excitation-pulse duration, we also observe Rabi oscillations on this transition (see Fig. \ref{fig:3}(e)). The PSP method increases the contrast by more than an order of magnitude to above 80\%.  Remaining deviations from the full contrast are explained by the residual ion temperature after Doppler cooling, which also contributes to the infidelity of the PSP process. From comparisons with the predicted behavior (see Supplemental Material \cite{appendixA}), we infer a mean motional state of $\bar{n}=18.5(33)$, corresponding to an ion temperature of $T=0.69(11)\, \rm{mK}$ that is close to the Doppler limit. We note that the repeated interrogation cycle during the PSP increases the preparation time. 

\begin{table*}[t]
\caption{\label{tab:1}
Measured hyperfine splittings $W_{FF'}$ and HFS constants obtained by including first-order (1st) and both first- and second-order (1st + 2nd) energy corrections. \m{The uncertainties of the HFS constants listed in the last column are primarily determined by the theoretical uncertainties of the second-order energy correction.}
}
\begin{ruledtabular}
\begin{tabular}{clll}
State & Hyperfine splitting (Hz) & HFS constant, 1st (Hz)
& HFS constant, 1st + 2nd (Hz)  \\ [0.5ex]
\hline \\ [-1.5ex]
${}^2\!S_{1/2}$ & $W_{23}=10\,491\,720\,234.7(4)$ & $A = -3\,497\,240\,078.23(13)$ & $A = -3\,497\,241\,700(600)$ \\ [0.8ex]
\hline \\[-1.5ex]
\multicolumn{1}{c}{}& $W_{12}=1\,007\,406\,257.7(14)$ & $A = -118\,257\,076.10(26)$ & $A = -118\,258\,070(120)$ \\ [0.5ex]
${}^2\!D_{3/2}$ & $W_{23}=788\,396\,819.9(14)$ & $B = \;\;\,963\,614\,049.4(12)$ &$B = \;\;\,963\,609\,800(570)$\\ [0.5ex]
\multicolumn{1}{c}{}& $W_{43}=297\,862\,646.5(14)$ & $C = -45.10(5)$ & $C = \;\;\,113(13)$ 
\end{tabular}
\end{ruledtabular}
\end{table*}

We employ the PSP method for state preparation and determine the excitation probability for frequencies around the resonance. From fitting the expected line shape to the recorded data, we obtain the center frequency. 
With this approach, we measure the isotope shift between $^{171}\rm{Yb}^+$ and $^{173}\rm{Yb}^+$ on the transition $^2\!{S}_{1/2} \rightarrow {}^2\!D_{3/2}$ for $F=0 \rightarrow F=2$ and $F=3 \rightarrow F=1$, respectively. 
Although we measure between $m_F=0$ levels, the relatively large applied magnetic fields cause second-order Zeeman shifts of a few kilohertz. From data obtained with different magnetic field magnitudes, we find the unperturbed isotope shift to be $6\,186\,981\,108.3(14)$ Hz \m{ with the uncertainty limited by statistics. The magnetic field magnitude is derived from spectroscopy of the $^2\!S_{1/2}$ hyperfine manifold as discussed below.} 

In addition to coherent optical spectroscopy, we perform MW spectroscopy using radiation referenced to a hydrogen maser to investigate the HFS of the $^2\!S_{1/2}$ and ${}^2\!D_{3/2}$ states.
For the former, with the ion prepared in the $^2\!S_{1/2}(F=3,m_F=0)$ state, we use 10-ms-long pulses to drive the hyperfine transition to the other hyperfine level ($F=2$) of the $^2\!S_{1/2}$ state at different magnetic fields. 
The resonant frequencies of the $m_F=0 \rightarrow m_F=0$ and $m_F=0 \rightarrow m_F=-1$ transitions are found from the corresponding excitation spectra. 
Substituting the measured frequencies into the Breit-Rabi formula \cite{Breit1931} yields the unperturbed HFS of $^2\!S_{1/2}$ and the magnetic field magnitude at the position of the ion. \m{While the primary source of uncertainty in magnetic field determination is the uncertainty of the magnetic g-factor $g_J=2.001(3)$ \cite{FAWCETT1991, Gossel2013}, the uncertainties of the frequencies deduced from the calibrated magnetic field magnitudes are dominated by statistics.}
This determination of the magnetic field magnitude is also used in the isotope shift measurement reported above and the HFS measurements of the ${}^2\!D_{3/2}$ state below. 

When measuring the HFS of ${}^2\!D_{3/2}$, we first prepare the ion in the $(F=1,m_F=0)$ level using PSP without the final pulse.
For measurements starting from other initial hyperfine levels, subsequent single or multiple MW $\pi$ pulses are used for preparation.
The limited coherence in the MW spectroscopy of the ${}^2\!D_{3/2}$ state is investigated using Rabi oscillations between the $F=1$ and $F=2$ levels.
The primary source of decoherence is the radiative lifetime of the ${}^2\!D_{3/2}$ state, which we find to be 56.3(35) ms (see Supplemental Material \cite{appendixA}). This result is compatible with 61.8(70) and 54.83(18) ms found for $\rm^{171}Yb^+$ $(F=2)$ \cite{Schacht2015} and $\rm^{174}Yb^+$ \cite{Shao2023}, respectively.
The hyperfine splittings of ${}^2\!D_{3/2}$ are derived from spectra obtained with 10-ms-long MW pulses between $m_F=0$ components with an extrapolation to the zero magnetic field.

The results of the HFS measurements are presented in Table \ref{tab:1} in the form of hyperfine splitting and HFS constants \cite{Schwartz1955}. 
Taking the hyperfine interaction to first order in perturbation theory, the measured splittings $W_{FF'}$ can be expressed as linear combinations of the HFS constants, which are then amenable to analytical determination. Accounting for both first- and second-order energy corrections within perturbation theory, however, requires a more comprehensive theoretical treatment that combines analytical derivations with numerical computations (see Ref.~\cite{Xiao2020} and Supplemental Material \cite{appendixA} for more details). 
Such a higher-order approach is essential for achieving the accuracy demanded by high-resolution spectroscopy.
\m{As seen from Table I, inclusion of the second-order energy corrections leads to contributions that are approximately 4 orders of magnitude larger than the experimental uncertainties to the HFS constants A and B.}


The $^2\!S_{1/2}$ state splitting is 4.9(4) Hz smaller than that found in the only previous measurement, which was carried out with trapped ions cooled by buffer gas \cite{Münch1987}. \m{A comparable discrepancy of 5.8(14) Hz has been observed between laser-cooled \cite{Han2024,Phoon2014} and buffer-gas-cooled \cite{Blatt1983} trapped ions for the measurement of the 12.6 GHz $^2\!S_{1/2}$ state splitting in $^{171}\rm{Yb}^+$. Because the discrepancies are similar in scale to the splittings, we speculate that they result from an erroneous frequency standard used in the earlier measurements with buffer-gas-cooled ions \cite{Blatt1983,Münch1987}.}

\begin{figure}[htbp]
\centering
\includegraphics[trim={0cm 0cm 0cm 0cm},clip,width=0.48\textwidth]{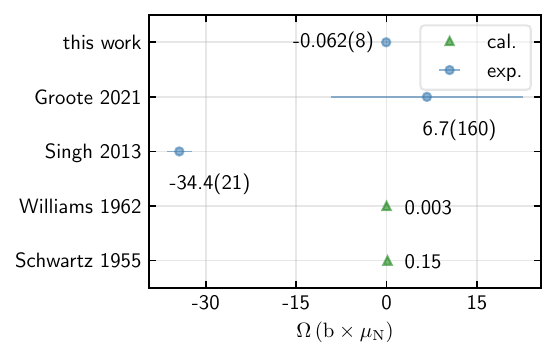}
\caption{The nuclear magnetic octupole moment $\Omega$ of $^{173}$Yb. The numbers beside the data points represent the corresponding values. Calculations (cal.) based on a single-particle model (Schwartz) \cite{Schwartz1955, Xiao2020} and including nuclear deformation (Williams) \cite{Williams1962} are shown as triangles; experimental results (exp.) of Singh \cite{Singh2013} and Groote \cite{Groote2021}  from HFS spectroscopy of the $^{3}\!P_2$ state in neutral $^{173}\rm{Yb}$ atoms are shown with results obtained in this work as circles.}
\label{fig:4}
\end{figure}

The size of the nuclear magnetic octupole moment $\Omega$ in $^{173}\rm{}Yb$ has been a subject of scientific debate \cite{Singh2013,Xiao2020} in recent years (see Fig. \ref{fig:4}). 
Using the hyperfine constant $C=113(13)\,{\rm Hz}$ deduced from the measured HFS and 
\m{the value of $C/\Omega = -1820(73)\,{\rm Hz}/({\rm b} \times \mu_N)$ calculated in the present work},
where $\rm{}b$ is barn and $\mu_N$ is the nuclear magneton, we obtain $\Omega = -0.062(8)$\,($\rm{b} \times\mu_N$).
We note that $\Omega$ is very sensitive to the second-order energy corrections \cite{Schwartz1955}. If we disregard them and use $C=-45.10(5)\,{\rm Hz}$, we arrive at $\Omega = 0.0248 (10) \,({\rm b} \times \mu_N)$, a value that differs even in sign from the result produced by the more elaborate and accurate treatment. 

Calculations of low-lying energy levels, HFS constants, and second-order energy corrections were carried out in the framework of a linearized coupled-cluster single double method~\cite{BlaJohSap91} and also using the AMBiT package~\cite{KAHL2019}. Calculation details, a comparison of the theoretical and experimental results, and a discussion of theoretical uncertainties are given in Supplemental Material \cite{appendixA}.

As shown in Fig.~\ref{fig:4}, the value of $\Omega$ derived from our measurement agrees with the result of \citet{Groote2021}, but differs substantially from the value given by \citet{Singh2013}.
Compared to these two measurements based on HFS spectroscopy of the $^{3}\!P_2$ state in $^{173}\rm{Yb}$, we improved the uncertainty in $\Omega$ by more than 2 orders of magnitude.
Our result indicates that $^{173}\rm{Yb}$ has a nuclear octupole moment in the range expected by nuclear theory \cite{Schwartz1955, Williams1962, Xiao2020}.

Our data from the HFS measurements can also be employed to investigate the differential hyperfine anomaly (DHA) \cite{Bohr1950, Roberts2021} between $^{171}\rm{Yb}$ and $^{173}\rm{Yb}$, which is denoted as ${}^{171}\Delta^{173}$ and defined as
\begin{equation}
A^{171}/A^{173}=g^{171}_I/g^{173}_I(1+{}^{171}\Delta^{173})\,, \nonumber
\end{equation}
where $g_I$ is the nuclear g-factor and $A$ is the hyperfine constant. Combining our result with the HFS of $^{171}\rm{Yb}^+$ measured in \cite{Han2024} ($^2\!S_{1/2}$) and \cite{Tamm2014} ($^2\!D_{3/2}$) results in 
\begin{align}
&{}^{171}\Delta^{173}(^2\!S_{1/2})=-0.65(8)\% \, , \nonumber
\\ &{}^{171}\Delta^{173}(^2\!D_{3/2})=-0.41(8)\% \, . \nonumber
\end{align}

For this calculation, we used the g-factors $g^{171}_I = 0.9898(8)$ and $g^{173}_I = -0.272008(12)$ deduced from the nuclear magnetic dipole moments measured with nuclear magnetic resonance (NMR) spectroscopy \cite{stone2014}.
The uncertainties of the DHA values are mostly determined by the uncertainty of $g^{171}_I$. In contrast to the octupole moment $\Omega$, the hyperfine anomalies for the $^2\!S_{1/2}$ and $^2\!D_{3/2}$ states have a low sensitivity to second-order energy corrections.
Including these corrections exerts no influence on the extracted anomalies at this accuracy.

The DHA is a measure of the deviation of the HFS from the case of a pointlike nucleus, as is observed in NMR experiments. Since the $S$ state electron wave function has the largest overlap with the nucleus, the largest DHA is expected to be found for the $S$ state. Our results agree with this expectation,
$|{}^{171}\Delta^{173}(^2\!S_{1/2})| > |{}^{171}\Delta^{173}(^2\!D_{3/2})|$, but also indicate a significant DHA for the $^2\!D_{3/2}$ state.
Our result for DHA of the $^2\!S_{1/2}$ state is in good agreement with a recent theoretical prediction of -$0.618\%$ \cite{Roberts2021}, but differs from the only previous experimental result of -$0.425(5)\%$ \cite{Münch1987}.

In conclusion, we have developed a laser cooling scheme compatible with ${}^2\!D_{3/2}$ state detection for trapped $^{173}$Yb$^+$ ions. Using a single trapped ion laser cooled close to the Doppler limit, the previously unobserved E2 transition $^2\!S_{1/2}(F=3) \rightarrow {}^2\!D_{3/2}(F=1)$ has been excited coherently, and the isotope shift between $^{171}\rm{Yb}^+$ and $^{173}\rm{Yb}^+$ on this transition has been measured. We have implemented state preparation of a specific Zeeman state using also this E2 transition. HFS of the ground and excited states of the observed E2 transition have been measured to extract the nuclear magnetic octupole moment and the hyperfine anomaly of $^{173}\rm{Yb}$. \m{The E2 transition that we have observed is a promising candidate for the study of nuclear-spin-dependent PNC \cite{Dzuba2011, Fortson1993},} which provides some of the most constraining low-energy tests of electroweak theory \cite{Robert2015, Safronova2018}.

The investigated $^2\!S_{1/2}(F=3) \rightarrow {}^2\!D_{3/2}(F=1)$ transition in $^{173}\rm{Yb}^+$ appears to be a suitable alternative to the $^2\!S_{1/2}(F=0,m_F=0) \rightarrow {}^2\!D_{3/2}(F=2,m_F=0)$ transition used for high-performance optical clocks based on the $^{171}\rm{Yb}^+$ isotope \cite{Tamm2014, STUHLER2021}.
Therefore, this work also establishes the foundation for the construction of optical clocks based on $^{173}\rm{Yb}^+$.

We gratefully acknowledge Wesley C. Campbell and Thomas Dellaert for discussion, Burghard Lipphardt and Andreas Hoppmann for experimental support, William J. Eckner for carefully reading the manuscript, and the Si-cavity team for providing a stable optical laser reference.  This work was supported by the Max Planck-RIKEN-PTB Center for Time, Constants and Fundamental Symmetries, by the Deutsche Forschungsgemeinschaft (DFG, German Research Foundation) under SFB 1227 DQ-mat–Project-ID 274200144–within project B02, and under Germany’s Excellence Strategy – EXC-2123 QuantumFrontiers – 390837967, and by the project 23IEM03 HIOC. The project (23IEM03 HIOC) has received funding from the European Partnership on Metrology, co-financed from the European Union’s Horizon Europe Research and Innovation Programme and by the Participating States. The theoretical work has been supported in part by the US Office of Naval Research Grant N000142512105, by the European Research Council (ERC) under the Horizon 2020 Research and Innovation Program of the European Union (Grant Agreement No. 856415), and by the Deutsche Forschungsgemeinschaft (DFG, German Research Foundation) -- project 544815538.

\paragraph*{Data availability---} 
\hspace{-1.5em} The data that support the findings of this article are openly available \cite{data}.

%

\clearpage
\clearpage


\section*{Supplementary material}

# Supplemental Material

## SKETCH OF THE EXPERIMENTAL SETUP

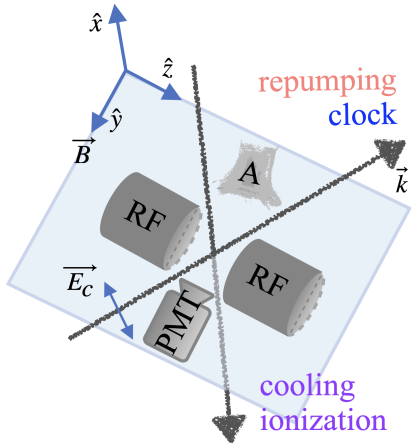

FIG. 1. A sketch of the experimental setup. Laser beams are focused into the center of the Paul trap (RF). The applied magnetic field, $\vec{B}$, is always perpendicular to the trap axis. A PMT is used for fluorescence detection. Antennas (A) are employed to drive MW transitions. The polarization $\vec{E}_c$ and the wave vector $\vec{k}$ of the E2 clock laser beam lie in the plane where $\vec{B}$ locates. The angle between $\vec{k}$ and $\vec{B}$, as well as between $\vec{E}_c$ and $\vec{B}$, is about 45°.

## RADIATIVE LIFETIME OF THE $^2D_{3/2}$ STATE

We perform Rabi flops on the $^2D_{3/2}(F=1, m_F=0) \to {}^2D_{3/2}(F=2, m_F=0)$ transition. The result is plotted in Fig. 2 as the population of the $^2D_{3/2}(F=1, m_F=0)$ state versus the duration of the applied pulse. The observed decoherence is attributed to the finite lifetime of the $^2D_{3/2}$ state. To describe this kind of decoherence, we introduce an exponential factor into the Rabi flop function, namely, define

$$P(t) = \cos\left(\frac{\Omega(t+t_0)}{2}\right)^2 \exp\left(-\frac{t+t_0}{\tau}\right), \quad (1)$$

where $P$ and $\tau$ are the population and the lifetime of the $^2D_{3/2}(F=1, m_F=0)$ state, $\Omega$ is the Rabi frequency of the transition being driven, and $t_0$ is the time period during which the ion may decay spontaneously before the observation is started, e.g. when the mechanical shutters used to prevent light shift are closing. Fitting Eq. 1 to the measured population of the $^2D_{3/2}(F=1, m_F=0)$ state results in a lifetime of $\tau = 56.3(35)$ ms.

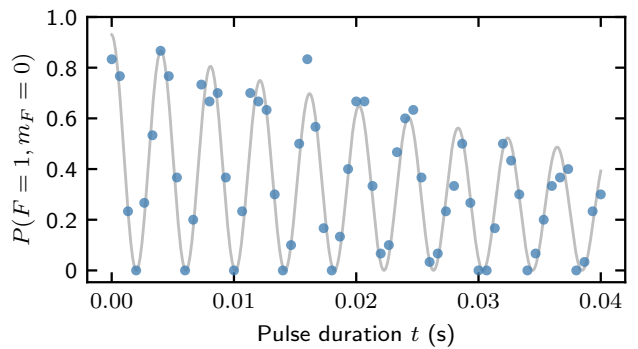

FIG. 2. Rabi flop on the transition $^2D_{3/2}(F=1, m_F=0) \to {}^2D_{3/2}(F=2, m_F=0)$. A Rabi flop function with exponential decay (line) is fitted to the data (points).

## RABI FLOP WITH DECOHERENCE INDUCED BY POPULATION OF MULTIPLE MOTIONAL STATES

For a trapped ion, a Rabi flop performed on a narrow transition, for example, the $^2S_{1/2}(F=3, m_F=0) \to {}^2D_{3/2}(F=1, m_F=0)$ transition in our study, shows decoherence when the resolved linewidth is smaller than the trapping frequency and the ion is not cooled to the motional ground state. Rabi flops with this kind of decoherence can be described by the hot Rabi flop function [1],

$$p(t) = \frac{1-\cos(\Omega t)+\eta^2 \bar{n}\Omega t \sin(\Omega t)}{2(1+(\eta^2 \bar{n}\Omega t)^2)}, \quad (2)$$

where $p$, $t$, $\Omega$, $\eta$, and $\bar{n}$ are the excitation probability, duration of the applied excitation pulse, Rabi frequency of the transition being driven, Lamb-Dicke parameter, and average motional state of the trapped ion, respectively. The Lamb-Dicke parameter $\eta$ that describes the fraction of the quadrupole component of the potential created by the applied voltage is defined for the endcap trap used in this study [2] as

$$\eta = \frac{2\pi}{\lambda}\sqrt{\frac{\hbar}{2m\omega_z(1-\frac{1}{2}\sin^2\phi)}}, \quad (3)$$

where $\lambda$, $\hbar$, $m$, $\omega_z$, and $\phi$ are the wavelength of the excitation radiation, the reduced Planck constant, the mass of the trapped ion, the trapping frequency along the axial trap axis, and the angle between the excitation radiation and the axial trap axis. In our case, $\eta = 0.084(7)$, where the uncertainty is inherited dominantly from the uncertainty of $\phi$.

Fitting the hot Rabi function to the data from the Rabi flop experiment, which is the solid curve in Fig.

2(e) in the main text, yields the mean motional state $\bar{n} = 18.5(3.3)$ of the ion.

## METHOD OF CALCULATION

We consider $Yb^+$ as a univalent ion with a core $[1s^2, 2s^2, ..., 4f^{14}]$ and a valence electron above it. The basis set is constructed in $V^{N-1}$ approximation in the framework of the Dirac-Hartree-Fock (DHF) approach. The initial DHF self-consistency procedure included the Breit interaction and was performed for the core electrons. Then, the $6s$, $6p$, and $5d$ orbitals were constructed in the frozen core potential. The remaining virtual orbitals were formed using a recurrent procedure described in Refs. [3, 4] on a nonlinear grid with 500 points. The last point of the radial grid was at 60 a.u. The basis set includes partial waves with the orbital quantum number up to $l = 5$ and orbitals with a principal quantum number $n$ up to 28.

The quantum electrodynamic (QED) corrections were also taken into account. For calculations, we applied a linearized coupled-cluster single double (LCCSD) method, solving the coupled-cluster equations in a basis set consisting of single-particle states.

## ENERGY LEVELS AND HYPERFINE STRUCTURE CONSTANTS

The numerical results of calculating the energies and hyperfine structure (HFS) constants of the $ns\,^2S_{1/2}$ and $5d\,^2D_{3/2,5/2}$ states, obtained in the framework of the LCCSD method, are presented in Table I. In calculating the HFS constants $A$ and $B$ we used the nuclear magnetic dipole moment $\mu = -0.68$, expressed in the nuclear magnetons $\mu_N$, and the electric quadrupole moment $Q = 2.80$ b.

## HYPERFINE STRUCTURE OF THE $6s\,^2S_{1/2}$ AND $5d\,^2D_{3/2}$ STATES

The hyperfine splittings of the $6s\,^2S_{1/2}$ and $5d\,^2D_{3/2}$ states were measured in this work. The first-order energy correction due to hyperfine interaction can be expressed as a linear combination of the HFS constants [7]. Using the experimental results and including the first- and second-order energy corrections [7] into consideration, we can find the HFS constants $A$, $B$, and $C$.

### The HFS constant $A$ of the $6s\,^2S_{1/2}$ state

The nuclear spin for $^{173}$Yb is $I = 5/2$. Accounting for the first- and second-order energy corrections, we find the

TABLE I. The theoretical and experimental [5] excitation energies of the $5d\,^2D_{3/2,5/2}$ and $(7-9)s\,^2S_{1/2}$ states are presented in the upper part. The HFS constants $A$, $B$, and $C/\Omega$ are listed and compared with the experimental values.

| | | Theory | Experim. |
|---|---|---|---|
| Energy (cm$^{-1}$) | $5d\,^2D_{3/2}$ | 22837 | 22961 |
| | $5d\,^2D_{5/2}$ | 24262 | 24333 |
| | $7s\,^2S_{1/2}$ | 54760 | 54304 |
| | $8s\,^2S_{1/2}$ | 73568 | 73040 |
| | $9s\,^2S_{1/2}$ | 83009 | |
| $A$ (MHz) | $6s\,^2S_{1/2}$ | -3646 | -3497[a] |
| | $5d\,^2D_{3/2}$ | -123 | -118[a] |
| | $5d\,^2D_{5/2}$ | 17.1 | 17.5[b] |
| $B$ (MHz) | $5d\,^2D_{3/2}$ | 990 | 964[a] |
| | $5d\,^2D_{5/2}$ | 1243 | |
| $C/\Omega$ (MHz/(b × $\mu_N$)) | $5d\,^2D_{3/2}$ | -0.001820 | |

[a] This work. The numbers are rounded to whole numbers. Exact values are given in the main text,
[b] Ref. [6]. Recalculated from $A$ measured for $^{171}$Yb$^+$.

energy shifts $W_F$ ($F = 2, 3$) for the $6s\,^2S_{1/2}$ state,

$$W_2 = -\frac{7}{4}A + \frac{7}{180}\eta$$
$$W_3 = \frac{5}{4}A + \frac{5}{252}\eta, \quad (4)$$

where $\eta$ is given by the expression

$$\eta \approx \frac{42}{5}\mu^2 \sum_{n \geq 7} \frac{\langle 6s\,^2S_{1/2}||T_1||ns\,^2S_{1/2}\rangle^2}{E(6s) - E(ns)}, \quad (5)$$

where $T_1$ is the magnetic-dipole hyperfine operator acting in the electron space, $E(ns)$ is the energy of the $ns\,^2S_{1/2}$ state, and $n$ is the principal quantum number.

Then, we have for the difference $W_{23} \equiv W_2 - W_3$,

$$W_{23} = -3A + \frac{2}{105}\eta. \quad (6)$$

To find the HFS constant $A$, we need to calculate $\eta$ and determine its uncertainty. As seen in Table I, the theoretical energies of the $7s$ and $8s$ states agree with the experimental data to an accuracy better than 1%. The difference between theory and experiment for $A(6s)$ is $\sim 4\%$ and, respectively, the uncertainty of the diagonal reduced matrix element (ME) $\langle 6s\,^2S_{1/2}||T_1||6s\,^2S_{1/2}\rangle$ is also $\sim 4\%$. We assume that the non-diagonal MEs $\langle 6s\,^2S_{1/2}||T_1||ns\,^2S_{1/2}\rangle$ are calculated with a similar accuracy.

An accurate calculation of $\eta$ is difficult due to the essential contribution of intermediate $ns$ states with $n \geq 7$.

To determine their role, we carried out three calculations: (i) including only the intermediate state $7s$ in Eq. (5), (ii) including two intermediate states $7s$ and $8s$, and (iii) including three intermediate states $7s$, $8s$, and $9s$.

Using the following values of the reduced MEs,

$$\langle 6s\,^2S_{1/2}||T_1||7s\,^2S_{1/2}\rangle \approx 8577,$$
$$\langle 6s\,^2S_{1/2}||T_1||8s\,^2S_{1/2}\rangle \approx 5602,$$
$$\langle 6s\,^2S_{1/2}||T_1||9s\,^2S_{1/2}\rangle \approx 4533, \qquad (7)$$

expressed in (MHz $\times\,\mu_N$), we obtain for $\eta$ (in MHz)

$$\eta \approx -0.175,\ \text{only } 7s \text{ is included},$$
$$\eta \approx -0.231,\ 7-8s \text{ are included},$$
$$\eta \approx -0.263,\ 7-9s \text{ are included}. \qquad (8)$$

Thus, the inclusion of the $8s$ and $9s$ intermediate states changed $\eta$ by 0.09 MHz. Considering this value as an uncertainty of $\eta$, we arrive at

$$\eta = -0.26(9)\ \text{MHz}.$$

Substituting $\eta$ and the experimentally measured $W_{23} = 10\,491.7202347(4)$ MHz in Eq. (6), we find the value of the HFS constant $A$ for the $6s\,^2S_{1/2}$ state,

$$A = -3497.2417(6)\ \text{MHz}. \qquad (9)$$

### The HFS constants of the $5d\,^2D_{3/2}$ state

We start from equations for the energy shifts $W_F$ ($F$=1–4) for the $5d\,^2D_{3/2}$ state, obtained with inclusion of the first- and second-order energy corrections. In calculating the second-order energy corrections, we account for the magnetic-dipole and electric-quadrupole hyperfine interaction and disregard the magnetic-octupole interaction. Then,

$$W_1 \approx -\frac{21}{4}A + \frac{7}{10}B - \frac{42}{5}C + a_1\eta_{\mu\mu} + b_1\eta_{\mu Q} + c_1\eta_{QQ},$$
$$W_2 \approx -\frac{13}{4}A - \frac{1}{10}B + \frac{54}{5}C + a_2\eta_{\mu\mu} + b_2\eta_{\mu Q} + c_2\eta_{QQ},$$
$$W_3 \approx -\frac{1}{4}A - \frac{11}{20}B - \frac{27}{5}C + a_3\eta_{\mu\mu} + b_3\eta_{\mu Q} + c_3\eta_{QQ},$$
$$W_4 \approx \frac{15}{4}A + \frac{1}{4}B + C + a_4\eta_{\mu\mu} + b_4\eta_{\mu Q} + c_4\eta_{QQ}. \qquad (10)$$

Here $A$, $B$, and $C$ are the magnetic-dipole, electric-quadrupole, and magnetic-octupole HFS constants of the $5d\,^2D_{3/2}$ state. The coefficients, $a_F$, $b_F$, and $c_F$ are the angular factors, given by the following formulas [7]

$$a_F = \begin{Bmatrix} I & J & F \\ J' & I & 1 \end{Bmatrix}^2,$$
$$b_F = \begin{Bmatrix} I & J & F \\ J' & I & 1 \end{Bmatrix}\begin{Bmatrix} I & J & F \\ J' & I & 2 \end{Bmatrix},$$
$$c_F = \begin{Bmatrix} I & J & F \\ J' & I & 2 \end{Bmatrix}^2. \qquad (11)$$

In the sum over intermediate states, we restrict ourselves to including only the $5d\,^2D_{5/2}$ state. Then, the quantities $\eta_{\mu\mu}$, $\eta_{\mu Q}$, and $\eta_{QQ}$ can be written as

$$\eta_{\mu\mu} = \frac{42}{5}\mu^2 \frac{\langle 5d\,^2D_{3/2}||T_1||5d\,^2D_{5/2}\rangle^2}{E(^2D_{3/2}) - E(^2D_{5/2})},$$
$$\eta_{\mu Q} = \frac{42\sqrt{2}}{5}\mu Q$$
$$\times \frac{\langle 5d\,^2D_{3/2}||T_1||5d\,^2D_{5/2}\rangle\langle 5d\,^2D_{3/2}||T_2||5d\,^2D_{5/2}\rangle}{E(^2D_{3/2}) - E(^2D_{3/5})},$$
$$\eta_{QQ} = \frac{21}{5}Q^2 \frac{\langle 5d\,^2D_{3/2}||T_2||5d\,^2D_{5/2}\rangle^2}{E(^2D_{3/2}) - E(^2D_{5/2})}, \qquad (12)$$

where $T_1$ and $T_2$ are the magnetic-dipole and electric-quadrupole hyperfine operators acting in the electron space. The MEs found in the framework of the LCCSD method are

$$\langle 5d\,^2D_{3/2}||T_1||5d\,^2D_{5/2}\rangle \approx -1733\ \text{MHz} \times \mu_N,$$
$$\langle 5d\,^2D_{3/2}||T_2||5d\,^2D_{5/2}\rangle \approx -453\ \text{MHz/b}. \qquad (13)$$

The theoretical and experimental values of the HFS constants $A(5d\,^2D_{3/2})$ and $B(5d\,^2D_{3/2})$, presented in Table I, differ by less than 4%. Assuming that the nondiagonal MEs $\langle 5d\,^2D_{3/2}||T_{1,2}||5d\,^2D_{5/2}\rangle$ are calculated

with a similar accuracy, we assign a 4% uncertainty to them. Respectively, the uncertainties of $\eta_{\mu\mu}$, $\eta_{\mu Q}$, and $\eta_{QQ}$ are $\sim 8\%$.

Using $W_F$, given by Eq. (10), we can write equations for $W_{12}$, $W_{23}$, and $W_{43}$ (where $W_{ik} \equiv W_i - W_k$). These quantities are measured at the experiment. Solving the system of three linear equations, we find the HFS constants $A$, $B$, and $C$ and determine their uncertainties,

$$A = -118.25807(12) \text{ MHz},$$
$$B = 963.60980(57) \text{ MHz},$$
$$C = 113(13) \text{ Hz.} \qquad (14)$$

### The octupole moment $\Omega$

Knowing the HFS constant $C$ and using the theoretical value $C/\Omega \approx -1820(73)$ Hz/(b $\times \mu_N$), presented in Table I, we can find the octupole moment $\Omega$.

To determine its uncertainty, we note that for calculating $C/\Omega$ we used the same wave function of the $5d\,^2D_{3/2}$ state that we used for calculating $A(5d\,^2D_{3/2})$ and $B(5d\,^2D_{3/2})$. The quality of the wave function was additionally tested by comparing the theoretical and experimental energies. Based on this, we assume that the uncertainty of $C/\Omega$ is comparable to the uncertainties of the HFS constants $A$ and $B$, i.e., $\sim 4\%$.

Finding the fractional uncertainty $\delta\Omega$ as

$$\delta\Omega = \sqrt{(\delta C)^2 + (\delta(C/\Omega))^2} \approx 0.122,$$

we obtain the octupole moment $\Omega$ to be

$$\Omega = -0.062(8) \ (b \times \mu_N). \qquad (15)$$



\end{document}